\input{aipcheck}

\documentclass[
    ,final            
  ]
  {aipproc}

\layoutstyle{6x9}

\newcommand{\beq}{\begin{equation}}
\newcommand{\eeq}{\end{equation}}

\begin{document}

\title{Dual string from lattice Yang-Mills theory}


\author{V.~I. Zakharov}{
  address={Max-Planck Institut f\"ur Physik \\
Werner-Heisenberg Institut \\ 
F\"ohringer Ring 6, 80805, Munich\\ 
E-mail: xxz@mppmu.mpg.de}
}



\begin{abstract}
We review properties of lower-dimension vacuum defects
observed in lattice simulations of SU(2) Yang-Mills theories. 
One- and two-dimensional defects are  associated with ultraviolet
divergent action. The action is  the same divergent as in perturbation theory but
the fluctuations extend  over submanifolds of the whole 4d space.
The  action is self tuned to a divergent entropy and
the 2d defects can be thought of as dual strings populated with particles.
The newly emerging 3d defects are closely related to the confinement mechanism.
Namely, there is a kind of holography
so that information on the confinement is encoded in a 3d submanifold.
We introduce an  SU(2) invariant classification scheme which 
allows
for a unified description of $d=1,2,3$ defects. The scheme
fits known data and predicts that 3d defects are related
to chiral symmetry breaking. Relation to stochastic vacuum model is 
briefly discussed as well.
\end{abstract}

\maketitle

\section{Introduction}

Studies of the confinement mechanism have become since long a prerogative
of the lattice simulations, for a recent review see \cite{greensite}.
The continuum theory provided in fact little guidance for 
search of the confinement mechanism. Equations which one borrows from
the continuum physics refer  mostly to U(1) Higgs models
or instantons, see, e.g., \cite{engelhardt}. However, these hints from 
the continuum theory could be used at a qualitative level at best.

Painstaking analysis of the lattice simulations did allow to extract
vacuum  fluctuations which are actually responsible for the confinement.
These are so called monopoles and central vortices, for review see,
e.g., \cite{suzuki} and \cite{greensite,langfeld}. By construction,
monopoles are infinitely thin closed trajectories while the
central vortices are infinitely thin closed 2d surfaces \footnote{
both trajectories and surfaces are defined actually on the dual lattice.}. 
Separation of the two types of the defects is actually superficial.
Rather, one observes vortices populated with monopoles.
Monopoles live on 2d surfaces, not in the whole 4d space
and there can be no vortices without monopoles,
\cite{gubarev,kovalenko}.                

Infinitely thin (with size of the lattice spacing $a$), percolating 
trajectories and surfaces look very different from,
say, instantons which are bulky fields, with size of order 
$\Lambda_{QCD}^{-1}$. Thus, one is tempted to say that lattice simulations 
uncovered existence of objects of lower dimensions in the vacuum 
state of Yang-Mills theories. However, a prevailing viewpoint, for
a recent presentation see, e.g., \cite{engelhardt}, is that apparent
point-likeness of the monopoles and vortices is an artifact
of their definition and in fact they only mark some
bulky field fluctuations.

It is only rather recently that it was understood that 
the monopoles and vortices might still be physical lower-dimensional
defects. The basic observation which brings about such a conclusion
is the ultraviolet divergence in the corresponding {\it non-Abelian}
action \footnote{It is worth emphasizing that the non-perturbative ultraviolet
divergent fields are no more divergent than perturbation theory,
for details see \cite{match}.} associated with the monopoles
\cite{anatomy} and vortices \cite{gubarev}.
The power of the ultraviolet divergence in the action is the same as for
pointlike particles and infinitely thin strings, respectively.
To explain the survival of the monopoles and vortices on the $\Lambda_{QCD}$ scale --
despite of their ultraviolet divergent action-- one is forced to postulate \cite{hierarchy}
self-tuning of the ultraviolet divergent action and of ultraviolet
divergent entropy \footnote{Visible entropy of the 2d and 1d defects explodes 
exponentially with the lattice spacing $a\to 0$ \cite{kovalenko}.}.  
Moreover, the d=2 defects appear to be nothing 
else but dual strings with excitations
of scalar field living on them.

The ultraviolet divergences in the action are the earliest 
evidence in favor of relevance of singular fields to
confinement. There exist further  observations 
\cite{horvath,volume,forcrand}
indicating
that lower dimensional defects are of physical significance
\footnote{Note that for lower-dimension defects to be relevant
the corresponding fields are to be singular.}.

When the lattice
studies were undertaken first, there was no theory of extended objects
at all. However, more recently the idea that strings are relevant to QCD
has become quite common. More specifically, one expects that if a dual
formulation of YM theories exists, it would be a string theory
\cite{maldacena} \footnote{Usually  one believes that it is only in
the limit of infinite number of colors, $N_c\to \infty$,
that one might find a dual formulation.}. 
 
Thus, there appears a possibility that the languages of lattice and
continuum theories would get unified again, this time in terms of
theory of extended objects.  A possible feedback from lattice studies to the
continuum theory is that topological excitations observed within a
`direct' formulation might become fundamental variables of the
dual formulation of the same theory,
see, e.g., \cite{savit}. Thus, if the strings are indeed
observed as excitations in lattice simulations of YM theories,
this is an
indication that there exists a dual formulation in terms of fundamental
strings \cite{hints}. 

Here we address a problem of reformulating 
some of the lattice results in
terms of the continuum theory.
The point is that many results, especially on the lattice strings,
are obtained originally in terms of so called projected fields,
see, e.g., \cite{greensite}.
We will discuss   a classification scheme of the defects in explicitly
SU(2) invariant terms. Also we will comment on possible relation
to the stochastic picture of the vacuum \cite{olesen,simonov}.

\section{Lattice strings}

We have reviewed recently the properties of the two-dimensional defects,
or lattice strings \cite{zakharov1} and will be brief here.

\subsection{Magnetic monopoles}

Theoretically, the most difficult point about the monopoles is their definition 
on the lattice. Monopoles are topological excitations of the compact
$U(1)$ \cite{polyakov2}. To define them in non-Abelian case one uses
projection of the original YM fields onto the `closest' Abelian configuration.
The physical idea behind considering the monopoles is that confinement
is mostly due to Abelian degree of freedom \cite{suzuki2}.

While the definition of the monopoles is not so transparent, many observed 
properties are beautiful and formulated in perfectly SU(2) invariant way.
Monopoles are observed as clusters of trajectories. Infinite, or percolating
cluster corresponds to the classical expectation value, $<\phi_M>$
of a magnetically charged field $\phi_M$. Short, or ultraviolet clusters correspond
to quantum fluctuations of the field $\phi_M$. The total length of the clusters
is trivially proportional to the total volume of the lattice, $V_4$: 
\beq
L_{tot}~=~4\rho_{tot}\cdot V_4~=~4(\rho_{perc}~+~\rho_{finite})\cdot V_4~~.
\end{equation}
According to the data \cite{boyko}:
\beq\label{total}
\rho_{tot}~\approx~1.6(fm)^{-3}~+~1.5(fm)^{-2}\cdot a^{-1}~~,
\end{equation}
where $a$ is the lattice spacing. The $a^{-1}$ term is entirely due to the finite clusters.
For the percolating cluster the density is a constant in the physical units.

One can translate (\ref{total}) into the 
standard filed theoretic language by observing that \cite{hierarchy}:
\beq
\langle ~|\phi_M|^2~\rangle~=~(const)a\cdot \rho_{tot}~~.
\end{equation}
Thus, we have
\beq\label{squared}
\langle ~|\phi_M|^2~\rangle~\sim~\Lambda_{QCD}^2 ~~.\end{equation}
Theoretically, the estimate (\ref{squared}) can be derived as a constraint implied by the
asymptotic freedom of YM theories \cite{match}.

\subsection{Point-like facet of the monopoles}

The  monopoles action diverges with $a\to 0$
and the power of the divergence is the same 
as for point-like particles \cite{anatomy}:
\beq\label{polymer}
S_{mon}~\equiv~M\cdot L_{mon}~,~~~
M(a) ~\approx~\ln 7\cdot a^{-1}~~,
\end{equation}
where $M(a)$ corresponds to the radiative mass and is found by measuring extra {\it non-Abelian} action
associated with the monopoles. We quote the data in a way
which allows for a straightforward theoretical interpretation.
Namely, in field theory (see, e.g., \cite{sym}) if one starts with the classical action
of a particle, $S=M\cdot L$ the propagating mass is not $M$ but:
\beq\label{propagating}
m^2_{prop}~=~{(const)\over a}\big(M(a)~-~{\ln 7\over a}\big)~~,
\end{equation}
where the  constants $const, \ln 7$ are of pure geometrical origin and depend on
the lattice used. In particular, $\ln 7$ corresponds to  
the hypercubic lattice. Note that in Euclidean space a physical mass of a point-like
particle can appear only as a result of tuning between divergent action and entropy.

Thus, the data (\ref{polymer}) correspond to a small monopole mass. Moreover,
data (\ref{total}) imply that globally monopoles live on a 2d surface. For
ordinary point-like particles $\rho_{tot}\sim a^{-3}$.

\subsection{Closed strings}

Closed surfaces are topological defects of the $Z_2$ gauge theory. 
In simulations of SU(2) theory these surfaces are defined in terms
of the closest $Z_2$ projection which replaces the original YM fields with $Z_{\mu}(x)=\pm 1$.
The central vortices are defined as unification of all the plaquettes on the
dual lattice which pierce negative plaquettes in the $Z_2$ projection,
for review see  \cite{greensite,langfeld}.

Two most striking properties of the central vortices is that their
total area scales in physical units, for review see \cite{greensite,langfeld} 
while non-Abelian action is
ultraviolet divergent \cite{gubarev}:
\beq\label{two}
A_{tot}~\approx~4~(fm)^{-2}V_4~,~~~S_{tot}~\approx~0.54{A_{tot}\over a^2}~~.
\end{equation}
Moreover, the excess of the action disappears on the plaquettes next to
those belonging to the vortex. In other words, the vortices are infinitely
thin, at least on the presently available lattices.

It is worth emphasizing that the properties (\ref{two})
amount to observing an elementary string. Indeed, the data on the total area
imply that the tension is of order $\Lambda_{QCD}^2$ while the ultraviolet
divergence in the action assumes vanishing thickness. 
The suppression due to the action is to be compensated by enhancement 
due to the entropy. Fine tuning of the entropy and action is a generic feature
of any consistent theory of an elementary string in Euclidean space.

Another striking feature of the lattice strings is that the monopole trajectories,
discussed in the preceding subsection, lie in fact on the central vortices
\cite{gubarev,kovalenko,giedt}.
Thus, the two types of defects merge with each other.

\section{Three dimensional domains}

The 3d defects are more recent than the strings and have been studied 
in less detail. Moreover, there are a few independent pieces of evidence
in favor of existence of 3d defects which are, in fact, not
necessarily related to each other.

\subsection{`Strong' potentials}

The central vortices are defined in terms of negative plaquettes in $Z_2$ projection.
In $Z_2$ projection links take on values $\pm 1$. Generically, the values $(+1)$ and $(-1)$
are the same frequent. One can, however, minimize the number of negative links. using
remaining $Z_2$ invariance. Physicswise, one fixes the gauge by localizing large
potentials on as a small number of links as possible. Since link values correspond
to potentials and are gauge dependent, one can wonder what is the objective meaning
of such minimization. The point is that minimizing, say, potential squared
one arrives at a gauge invariant quantity \cite{stodolsky}. Minimizing number
of negative links is a variation of such a procedure. 

And, indeed, one finds \cite{volume}
that volume of negative links scales as a physical 3d defect:
\beq\label{four}
V_3~=~c_3\Lambda_{QCD}\cdot~V_4~~.
\end{equation}
Note that by construction the volume is bound by the central vortices.
This volume can be called Dirac volume \cite{greensite}.
Eq (\ref{four}) then states that the minimal Dirac volume scales in physical
units or, alternatively, has a zero fractal dimension.

\subsection{Holography and confinement}

Relation of the volume discussed above to the confinement is revealed through a 
remarkable observation of the authors of Ref \cite{elia}.
One replaces the original link matrices $U_{\mu}(x)$ by
$\tilde{U}_{\mu}(x)$ 
where
\beq\label{modified}
\tilde{U}_{\mu}(x)~\equiv~U_{\mu}(x)\cdot Z_{\mu}(x)~,
\end{equation}
where $Z_{\mu}(x)$ is the projected value of the same link. Next, one evaluates the Wilson loop
and quark condensate $\langle\bar{q}q\rangle$ in terms of the modified 
links $\tilde{U}$. The result \cite{elia}  
is that both the confining potential and spontaneous breaking of
the chiral symmetry disappear.

Originally \cite{elia} the change (\ref{modified}) affected approximately half of the
total number of links. Now, we see that it is enough to perform the change (\ref{modified})
on a 3d submanifold  to kill the confinement and 
chiral symmetry breaking. In other words, substitution (\ref{modified}) is an ad hoc 
modification in the ultraviolet of the fields on a 3d volume plus pure
gauge transformations.
 Thus, we observe a 
kind of holography, with information on
the confinement being encoded on a submanifold of the whole space.

In more detail, consider a plane on which we will draw a Wilson line. 
Consider, furthermore, a particular configuration of the gauge fields 
generated with the standard SU(2) action. Determine then the 3d volume 
described in the preceding section. Intersections of this volume with
the plane considered are segments of 1d lines. Now, we can draw any Wilson line
on the plane. The statement is that the sign of the Wilson line can be 
determined by counting the number of intersections with segments of 1d defects.
It is a highly non-trivial observation, challenge to interpret. 
Note that there is no logical contradiction, though. Indeed, there are gauges where
the confining fields are soft, of order $A_{\mu}\sim\Lambda_{QCD}$. 
Apparently, one can use gauge
invariance to choose a gauge where the confining fields are of order $A_{\mu}\sim 1/a$
but occupy a 3d volume \footnote{Some considerations on possible relation
between gauge invariance and holography in the gravitational case
can be found in \cite{thooft3}}. 

\subsection{Chiral symmetry breaking}

There is a series of observations, not directly related to each other that indicate
relevance of some 3d defects to the spontaneous breaking of the chiral
symmetry \footnote{We are considering the quenched approximation.}:

(a) procedure of Ref \cite{forcrand} described above makes also
the quark condensate vanish:
\beq
\langle \bar{q}q\rangle _{\tilde{U}}~\approx~0
\end{equation}
Now, we know \cite{volume} that the change (\ref{modified}) affects not a finite part
of the 4d space but only a 3d submanifold.

(b) there is evidence in favor of long range topological structure in QCD vacuum
which is related to chiral symmetry breaking \cite{horvath}. The search process for 
the topological structure is formulated in terms of eigenfunctions of the Dirac operator
and explicitly gauge invariant
\footnote{Moreover, measurements \cite{horvath} refer to $SU(3)$ color group.}.

(c) One introduces the so called inverse participation ratio,
see in particular \cite{gattringer},
 defined in terms of 
eigenfunctions of the Dirac operator:
\beq
I~=~N\Sigma_{x}\rho_i^2(x)~~,
\end{equation}
where $N$ is the number of lattice sites $x,\rho_i(x)~=~\psi^{\dagger}_i\psi_i(x)$,
and $\psi_i(x)$ is the $i$-th normalized $\big(\Sigma_{x}\rho_i(x)=1\big)$  lowest
eigenvector of the Dirac operator. 

Dependence of the inverse participation ratio on
the lattice spacing $a$ was studied in Ref \cite{forcrand}.
The result is:
\beq\label{result}
\langle~I~\rangle~=~c_1~+~c_2\cdot a^{-\gamma}~~,
\end{equation}
with a non-vanishing exponent $\gamma$:
                   $$1~\le~\gamma~\le~2~~.$$ Note that the value $\gamma=1$
would correspond, in the limit $a\to 0$ to localization of the eigenfunctions on a
3d volume. It is worth emphasizing that the $a$ dependence observed refers to 
an explicitly gauge invariant quantity.

To summarize, there are indications that the chiral symmetry breaking
is determined by gauge fields living on a subspace. Since the confinement itself
also seems to be related to a 3d volume (see above), it is not clear whether 
we deal with a phenomenon specific for chiral symmetry breaking or with an
effect common to confinement.

\section{Classification scheme}

\subsection{Invariants}

There is no theory of the defects in the non-Abelian case. 
However, even in the absence of such a theory one can try 
to find a SU(2) invariant classification scheme.
Generically, the first example of such a scheme for monopoles was proposed
long time ago \cite{thooft4}.
In pure YM theory, there are no classical monopole solutions.
However, imagine that there exists a scalar field, vector in the color space $H^a,~a=1,2,3$.
Then one could fix the gauge by rotating vector $H^a$ to the third direction
at each point. This fixation of the gauge would fail however at the points where
\beq\label{three}
H^a~=~0~~.\end{equation}
Condition (\ref{three}) can be viewed as three equations defining   1d defects in the 4d space
which can be  identified
with monopole trajectories \cite{thooft4}. It is crucial that (\ref{three}) is SU(2) invariant.

For various reasons, this idea does not seem to work in the realistic case,
for review and references see \cite{gubarev1}. Rather,  monopoles are associated with singular
non-Abelian fields (see above). Let us try to adjust the classification scheme to this
set up \cite{zakharov1,gubarev1,gubarev2}.

Begin with YM theory in three dimensions and assume that monopoles
violate the Bianchi identities. If the Bianchi identities 
\beq
D\tilde{G}~=~0~,
\end{equation}
hold,
the potential $A$ can be expressed in terms of the field strength tensor, 
see, e.g., \cite{halpern}:
\beq\label{inversion}
A~=~{1\over g}(\partial \tilde{G}) \tilde{G}^{-1}~~.
\end{equation}
The inverse matrix exists unless the determinant constructed on the components
of $\tilde{G}$ vanishes. Denoting $\tilde{G}^a_{ik}\equiv\epsilon_{ikl}B^a_l$
we have, therefore, the following condition for the Bianchi identities to be violated:
\beq\label{determinant}
det(B^a_i)~\equiv~\epsilon_{ikl}\epsilon_{abc}B^a_iB^b_kB^c_l~=~0~~.
\end{equation}
Note that the condition (\ref{determinant}) is perfectly gauge and rotation
invariant. Moreover, it singles out a surface (or a line on the dual lattice)
while monopoles are usually 0d defects in the 3d case.

Let us now consider the 4d Euclidean case. The rotational
group in 4d splits into a product of two $O(3)$ groups, $O(4)=O(3)\times O(3)$.
The corresponding representations of the $O(3)$ groups are chiral gluon fields
$(H^a_i\pm E^a_i)$.
Looking for a generalization of (\ref{determinant}) we notice that there are
now two possibilities:
\beq
det{(E^a_i~+~H^a_i)}~=~0~,~~or ~~det{(E^a_i~-~H^a_i)}~=0~.
\end{equation}
Imposing either of them we specify a 3d defect. On this 3d submanifold
one can use as independent three fields of a certain chirality
but not of the opposite one. 
Thus, association of 3d defects with chiral symmetry breaking arises as
a consequence of the symmetry of the problem.

The boundary of these 3d defects is determined by conditions;
\beq\label{both}
det{(E^a_i~+~H^a_i)}~=~0~,~~and ~~det{(E^a_i~-~H^a_i)}~=0~.
\end{equation}
which determine 2d defects.
Moreover, if both conditions (\ref{both}) are satisfied, there is no inversion
of the Bianchi identities similar to (\ref{inversion}). 

Finally, zeros of a second order  of the determinant
would define 1d defects. They automatically fall onto the 2d defects as well.

\subsection{Classification scheme vs data}

The classification scheme proposed above is based on symmetry alone and is not unique.
But, nevertheless, let us try to identify the 2d and 1d defects 
arising within this scheme with the central vortices and monopoles.
There are a few quite remarkable confirmations of such an identification:

(a) the 2d defects are associated, according to the scheme, with
singular fields and, possibly, violations of the Bianchi identities
And, indeed, the central vortices carry a
singular action \cite{gubarev}. Moreover, monopoles live on the vortices, 
on one hand, and may well signify
violation of the Bianchi identities, on the other;

(b) non-Abelian fields associated with the 2d defects are aligned with the surface.
This is confirmed by the measurements, according to which the excess of the action
vanishes already on the plaquettes next to the central vortices \cite{gubarev};

(c) the monopole trajectories are predicted to lie on the central vortices,
in agreement with the data \cite{giedt,gubarev};

(d) `monopoles' appear to be Abelian fields since zero of second order of
the determinant constructed on three independent (within a 3d defect) fields
implies that there is only a single independent color vector. Thus, monopoles
can well be detected through the U(1) projection.

(e) on the other hand, the non-Abelian field of the monopoles is not spherically
symmetrical but rather aligned with the surface. This collimation
of the field was observed in measurements, \cite{giedt}.

It is worth emphasizing that all the properties (a) - (e) are gauge invariant.
Thus, the data so far do confirm that through projections one detects gauge invariant
objects.

Finally, the scheme predicts that breaking of the chiral symmetry is associated with
3d defects. The corresponding lattice data were summarized in the preceding section.

\section{Stochasticity}

In the continuum limit, association of the confining fields with lower-dimension
defects implies stochastic-type of correlators
\footnote{The material of this section is based
on discussions with M.I. Polikarpov.}. Indeed, the 3d volumes, e.g., are
`not visible' in the continuum limit. $a\to 0$. 
Denote by $\bar{A}$ the confining potential obtained in
the gauge minimizing the number of negative links (see above).
Then
\beq
\langle~ \bar{A}(x),~\bar{A}(y)~\rangle~=~\Lambda_{QCD}\cdot \Lambda_{UV}~ f_{sing}(x-y)~+(regular~terms)~,
\end{equation}
where
$$f_{sing}(0)~=~1, ~~f_{sing}(x\neq 0)=0~~.$$
The singular nature of the confining potential could explain observed dependence 
of the localization of zero modes on the lattice spacing, see above.

It is worth emphasizing, however, that reduction of the confining potential 
to the `white noise'
would be a great oversimplification
\footnote{Actually, the `white noise' would not confine.}. 
Indeed, the 3d nature of the domains assumes also non-trivial
correlators for the derivatives of the potential.
The issue deserves further consideration.

Consider now contribution of strings into an
explicitly gauge invariant correlator:
\beq
\langle G^2(x),~G^2(y)\rangle_{strings} ~=~(const) \Lambda^4_{QCD}
\Lambda_{UV}^4f_{sing}(x-y)~+ ~(const)
\Lambda_{QCD}^8~f_{phys}(x-y)~,
\end{equation} 
where
$f_{phys}$ depends on the physical mass scale.
Note that appearance of the extra factor $\Lambda_{QCD}^4$ in front of $f_{phys}$ is of
pure geometrical origin and reflects relative suppression of the 2d volumes compared
to a 4d volume. On the other hand, appearance of the ultraviolet cut off in a non-local
term would contradict the asymptotic freedom. It is one more example of consistency of
the lattice strings with the asymptotic freedom, see also \cite{match}.

Finally, for a stochastic model of the confinement (see, e.g., \cite{olesen,simonov})
it is the correlator of two non-Abelian fields connected by a `Dirac-string' operator,
$$\langle G^a_{\mu\nu}(x)\Phi_{ab}(x-y)G^b_{\mu\nu}\rangle~=~D(x-y)~~,$$
 which is crucial. The contribution of the string, discussed above, to this
correlator is of the form:
\beq
D_{string}(x-y)~=~(const)\cdot f_{sing}(x-y)\Lambda_{QCD}\cdot \Lambda_{UV}^2~.
\end{equation}
Moreover, using standard approximations of the stochastic model \footnote{
Using the minimal area spanned on the Wilson line is the most sensitive point,
difficult to justify theoretically \cite{olesen}.}
one     obtains for the string tension $\sigma$ determining the heavy quark potential at
large distances:
\beq
\sigma~\approx~\theta_{string}\Delta S_{string}~\approx~{1\over 2}\sigma_{exp}~~,
\end{equation}
where $\theta_{string}$ is the probability of a given plaquette to belong to the
lattice string, $\Delta S_{string}$ is the extra
action associated with a plaquette belonging to the string, $\sigma_{exp}$
is the value obtained in simulations.

It is interesting that the correlator $D(x-y)$ is singular in any case,
$$lim_{|x-y|\gg a}{D(x-y)}~\sim\exp(-c|x-y|/a)$$
because the Dirac string, $\Phi(x-y)$ is a color object and has infinite self energy
\footnote{This point was emphasized to me by Ph. de Forcrand.
The well known measurements of the correlator $D(x-y)$ \cite{digiacomo}
use smoothened, or cooled fields. We discuss here description in terms of the
original field  configurations. We assume also that pure perturbative fields
drop out from the string-tension evaluation.}. Thus, the singular nature of 
the confining fields, see (\ref{two})
is the only mechanism which can make the stochastic model 
relevant.

\section{Conclusions}

Physics of confinement might undergo quite a dramatic change soon.
There have been emerging data indicating relevance to confinement of lower-dimension
defects, or singular fields. Two-dimensional defects with divergent action and entropy, 
which 
selftuned to each other are naturally interpreted as the 
dual string, observed as a vacuum excitation.
The string  possesses many SU(2) invariant properties
but is detected through projections.
Other emerging phenomena, 
a kind of holography and localization of modes on a submanifold shrinking
to zero with $a\to 0$, are observed in explicitly SU(2) invariant terms.
The price is that the structure of the fields responsible for these observational phenomena
is less transparent.

\section*{Acknowledgments}

I am thankful to  W. Bardeen, M.N. Chernodub, A. DiGiacomo, A. Gorsky, F.V. Gubarev, I. Horvath,
 J. Greensite,  M.I. Polikarpov, A. Polyakov, L. Stodolsky, L. Susskind, T. Suzuki,
A. Vainshtein
for discussions.
  
This mini review is based on the talks presented at the conferences 
``Quark Confinement and the Hadron Spectrum VI'' (Village Tanka, September 2004)
and ``QCD and String Theory'' (Santa Barbara, November 2004). 

This research was supported in part 
by the National Science Foundation under Grant No. PHY99-07949 and 
by the Grant INTAS 00-00111.


\begin{thebibliography}{0}


\bibitem{greensite}
J. Greensite, {\it Prog. Part. Nucl. Phys.} {\bf 51} 1 (2003) 
(hep-lat/0301023).

\bibitem{engelhardt}
M. Engelhardt, {\it ``Generation of confinement and other 
nonperturbative effects by infrared gluonic degrees of freedom''},
 hep-lat/0409023.


\bibitem{suzuki}
 M.N. Chernodub, F.V. Gubarev, M.I. Polikarpov, 
A.I. Veselov,
{\it Progr. Theor. Phys. Suppl.} {\bf 131}, 309 (1998);\\
A. Di Giacomo, {\it Progr. Theor. Phys. Suppl.} {\bf 131},
161 (1998) (hep-lat/9802008);  \\
T. Suzuki, {\it Progr. Theor. Phys. Suppl.} {\bf 131},
633 (1998).

\bibitem{langfeld}
K. Langfeld, {\it et. al.}, 
{\it ``Vortex induced confinement and the IR properties
of Green functions''},  hep-lat/0209173. 



\bibitem{boyko}
V.G. Bornyakov, P.Yu. Boyko, M.I. Polikarpov, V.I. Zakharov,
{\it  Nucl. Phys.} {\bf B672}  222 (2003) (hep-lat/0305021);


\bibitem{gubarev}
F.V. Gubarev, et al., {\it Phys. Lett.} {\bf B574} (2003) 136,
(hep-lat/0212003)


\bibitem{kovalenko}
A.V. Kovalenko, M.I. Polikarpov, S.N. Syritsyn, V.I. Zakharov,
{\it ``Interplay of monopoles and P-vortices''}, hep-lat/0309032.

\bibitem{anatomy}
V.G. Bornyakov, et al., {\it Phys. Lett.} {\bf B537} (2002) 291.

\bibitem{hierarchy}
V.I. Zakharov, {\it ``Hidden mass hierarchy''}, (hep-ph/0202040);\\
M.N. Chernodub, V.I. Zakharov, {\it Nucl. Phys.}
{\bf B669}, 233 (2003) (hep-th/0211267);\\
V.I. Zakharov, {\it Usp. Fiz. Nauk} {\bf  47} 39 (2004).


\bibitem{match}
V.I. Zakharov, {\it ``Fine tuning in lattice SU(2) glyuodynamics versus continuum theory constraints''},
hep-ph/0306262;\\
V.I. Zakharov, {\it ``Nonperturbative match of ultraviolet renormalon''},
hep-ph/0309178.



\bibitem{horvath}
I. Horvath, et al.,
{\it  Phys. Rev.} {\bf D68} (2003) 114505 (hep-lat/0302009);\\
I. Horvath, {\it `` The analysis of space-time structure in QCD vacuum: 
localization vs global behaviour in local observables and Dirac eigenmodes''},
 hep-lat/0410046.


\bibitem{volume}
M.I. Polikarpov, S.N. Syritsyn, V.I. Zakharov,
{\it  ``A novel probe of the vacuum of the lattice gluodynamics''},
hep-lat/0402018;\\
A.V. Kovalenko, M.I. Polikarpov, S.N. Syritsyn, V.I. Zakharov,
{\it ``Three dimensional vacuum domains in four dimensional
SU(2) gluodynamics''},  hep-lat/0408014.

\bibitem{forcrand}
MILC Collaboration (C. Aubin et al.),
{\it ``The scaling dimension of low lying Dirac eigenmodes of the
topological charge density''}  hep-lat/0410024

\bibitem{maldacena}
J.M. Maldacena, {\it Adv. Theor. Math.} {\bf 2}, 231 (1998),
(hep-th/9711200);\\
A.M. Polyakov, {\it Int. J. Mod. Phys.} {\bf A14}, 645 (1999),
(hep-th/9809057).

\bibitem{savit}
R. Savit, {\it  Rev. Mod. Phys.} {\bf 52} (1980) 453.

\bibitem{hints}
V.I. Zakharov, 
{\it ``Hints on dual variables from lattice 
$SU(2)$ gluodynamics''}, hep-ph/0309301.

\bibitem{olesen}
J. Ambjorn, P. Olesen, {\it  Nucl. Phys.} {\bf B170} (1980) 60;\\
J.M. Cornwall, {\it Phys. Rev.} {\bf D26} (1982) 1453;\\
T.I. Belova, Yu.M. Makeenko, M.I. Polikarpov, A.I. Veselov,
{\it  Nucl. Phys.} {\bf B230} (1984) 473;\\
H.G. Dosch, {\it Phys. Lett.} {\bf B190} (1987) 177;\\
H.G. Dosch, Yu.A. Simonov, {\it Phys. Lett.} {\bf B205} (1988) 339.

\bibitem{simonov}
A. Di Giacomo, H.G. Dosch, V.I. Shevchenko, Yu.A. Simonov,
{\it  Phys. Rept.}, 372 (2002) (hep-ph/0007223).

\bibitem{zakharov1}
V.I. Zakharov, {\it ``Lower-dimansion vacuum defects in lattice Yang-Mills
theory''}, hep-ph/0410034.

\bibitem{polyakov2}
 A.M. Polyakov, {\it Phys. Lett.} {\bf  B59} (1975) 82.

\bibitem{suzuki2}
T. Suzuki, I. Yotsuyanagi, {\it Phys. Rev.} {\bf D42} (1990) 4257.


\bibitem{sym}
J. Ambjorn, {\it ``Quantization of geometry''}, hep-th/9411179.


\bibitem{giedt}
J. Ambjorn , J. Giedt, J. Greensite, {\it  JHEP}  0002:033, (2000),(hep-lat/9907021).


\bibitem{stodolsky}
F.V. Gubarev, L. Stodolsky, V.I. Zakharov, {\it Phys. Rev. Lett.} {\bf 86},
2220 (2001) (hep-ph/0010057).

\bibitem{elia}
Ph. de Forcrand, M. D'Elia, {\it  Phys. Rev. Lett.}
{\bf 82}, 458 (1999) (hep-lat/9901020).

\bibitem{thooft3}
G. 't Hooft, {\it ``The hidden information in the standard model''}, ITP-UU-02/68. 

\bibitem{gattringer}
Ch. Gattringer, et al., 
{\it  Nucl. Phys. } {\bf B617} (2001) 101,( hep-lat/0107016);\\
J. Gattnar, et al., {\it ``Center vortices and Dirac eigenmodes in SU(2) lattice gauge theory''}.
hep-lat/0412023.
 
\bibitem{thooft4}
G. t'Hooft, {\it Nucl. Phys.}, {\bf B190}, 455 (1981).

\bibitem{gubarev1}
F.V. Gubarev, V.I. Zakharov,
{\it ``Interpreting the lattice monopoles in the continuum terms''} 
hep-lat/0211033.

\bibitem{gubarev2}
F.V. Gubarev, V.I. Zakharov, {\it Int. J. Mod. Phys.} {\bf A17} (2002) 157
(hep-th/0004012).



\bibitem{halpern}
   M.B. Halpern,{\it Phys. Rev.} {\bf D19} (1979) 517;\\
O. Ganor, J. Sonnenschein,
{\it  Int. J. Mod. Phys.} {\bf A11} (1996)  5701,
(hep-th/9507036).

\bibitem{digiacomo}
A. Di Giacomo, H. Panagopoulos, Phys. Lett., B285 (1992) 133;\\ 
A. Di Giacomo, E. Meggiolaro, H. Panagopoulos, Nucl Phys. B483 (1997) 371.

\end{thebibliography}
\end{document}